\documentclass[10pt]{article}

\usepackage{graphicx}%
\usepackage{amsmath,amssymb,amsthm,upref,bm}%
\usepackage{labelfig}%
\usepackage{mathrsfs} 

\DeclareMathAlphabet{\varmathbb}{U}{pxsyb}{m}{n}

\newtheorem{corollary}{Corollary}%
\newtheorem{theorem}{Theorem}%

\newcommand{\D}{\mathrm{d}\kern0.2pt}%
\newcommand{\ii}{\kern0.05em\mathrm{i}\kern0.05em}%
\newcommand{\E}{\textrm{e}}%
\renewcommand{\vec}[1]{\bm{#1}}%
\newcommand{\RR}{\mathbb{R}}%
\newcommand{\CC}{\mathbb{C}}%

\begin{document}

\baselineskip=4.4mm

\makeatletter

\title{\bf The floating-body problem: \\ an integro-differential equation \\ without
irregular frequencies}

\author{Nikolay Kuznetsov}

\date{}

\maketitle

\vspace{-8mm}

\begin{center}
Laboratory for Mathematical Modelling of Wave Phenomena, \\ Institute for Problems
in Mechanical Engineering, Russian Academy of Sciences, \\ V.O., Bol'shoy pr. 61,
St. Petersburg 199178, Russian Federation \\ E-mail: nikolay.g.kuznetsov@gmail.com
\end{center}

\begin{abstract}
\noindent The linear boundary value problem under consideration describes
time-harmonic motion of water in a horizontal three-dimensional layer of constant
depth in the presence of an obstacle adjacent to the upper side of the layer
(floating body). This problem for a complex-valued harmonic function involves mixed
boundary conditions and a radiation condition at infinity. Under rather general
geometric assumptions the existence of a unique solution is proved for all values of
the nonnegative problem's parameter related to the frequency of oscillations. The
proof is based on the representation of solution as a sum of simple- and
double-layer potentials with densities distributed over the obstacle's surface, thus
reducing the problem to an indefinite integro-differential equation. The latter is
shown to be soluble for all continuous right-hand side terms for which purpose S.~G.
Krein's theorem about indefinite equations is used.
\end{abstract}

\setcounter{equation}{0}

\section{Introduction}

In his remarkable article \cite{J}, John investigated the floating-body problem
which describes time-harmonic motion of an inviscid, incompressible, heavy fluid,
say water, in the presence of an immersed surface-piercing body. Two questions need
to be addressed concerning this problem which is linear under the assumption that
oscillations of water are of small amplitude. The first one is the question of
uniqueness and John demonstrated that the problem has a unique solution for bodies
satisfying the geometrical condition (now, it is referred to as John's condition),
requiring the wetted parts of the body's surface be confined to vertical cylinders
whose generators go through the intersections of this surface and that of the water
at rest (see Fig.~1). Moreover, it was widely believed that the uniqueness theorem
is true for all geometries and frequencies and it would be only a matter of time
before a general uniqueness proof was obtained. Many partial uniqueness results that
have been obtained so far are reviewed in detail in \cite{LWW}.

The situation had changed in 1996, when M. McIver \cite{MM} constructed the first
example of a family of surface-piercing structures that all support the same
two-dimensional trapped mode in infinitely deep water; that is, there is no
uniqueness in the floating-body problem for any of these structures. Another family
of structures that support an axisymmetric trapped mode in deep water is a
straightforward modification of the first one (see \cite{MP}). It should be
emphasised that one finds rigorous proof of the existence of trapped modes only for
these two examples and their simple generalisations; see \cite{LWW}, \S\,4.1.

The present paper deals with the second question which concerns the existence of a
solution at all frequencies. (Of course, we assume that John's condition holds
because this guarantees the uniqueness at all frequencies which is essential for our
considerations.) A natural way to resolve this question is to seek a solution using
the potential theory techniques, but the standard method based on a single-layer
potential does not yield a solution for some sequence of frequencies known as
irregular (see \S\,1.3 for details). This is the reason to propose a new approach in
the framework of the potential theory. Its advantage is not only the absence of
irregular of frequencies, but also the fact that it involves only potentials whose
densities are distributed over the wetted surface of the floating body. A few
alternative methods are outlined in \S\,1.4. Prior to describing our approach,
statement of the problem is given along with some auxiliary results.

\begin{figure}
\begin{center}
  \SetLabels
  \L (0.96*0.66) {\small $x_2$}\\
  \L (0.84*0.65) {\small $F$}\\
  \L (0.517*0.95) {\small $y$}\\
  \L (0.487*0.575) {\small $x_1$}\\
  \L (0.56*0.59) {\small $F$}\\
  \L (0.36*0.34) {\small $\Gamma$}\\
  \L (0.69*0.43) {\small $\Gamma$}\\
  \L (0.05*0.65) {\small $F$}\\
  \L (0.81*0.75) {\small $D$}\\
  \L (0.114*0.75) {\small $D$}\\
  \L (0.55*0.25) {\small $\Omega$}\\
  \L (0.37*0.607) {\small $\mathcal{B}$}\\
  \L (0.3*0.46) {\small $\Xi$}\\
  \L (0.71*0.56) {\small $\Xi$}\\
  \L (0.06*0.97) {\small $g$}\\
  \L (0.85*0.085) {\small $B$}\\
  \L (0.02*0.8) \includegraphics[width=3mm]{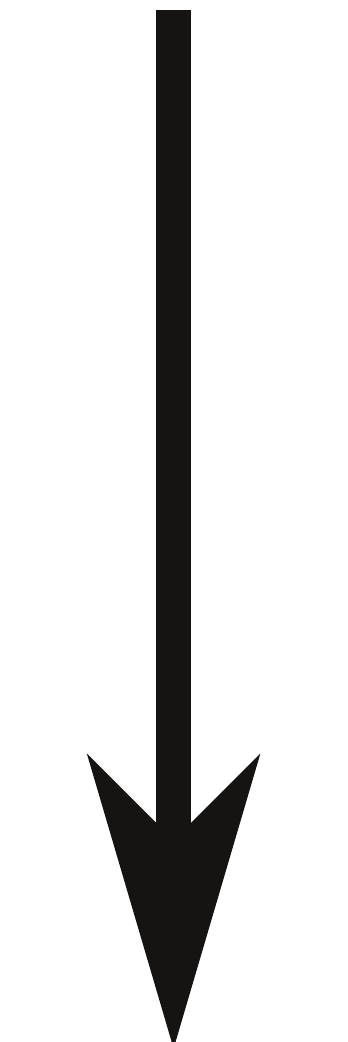}\\
  \endSetLabels
  \leavevmode
\strut\AffixLabels{\includegraphics[width=58mm]{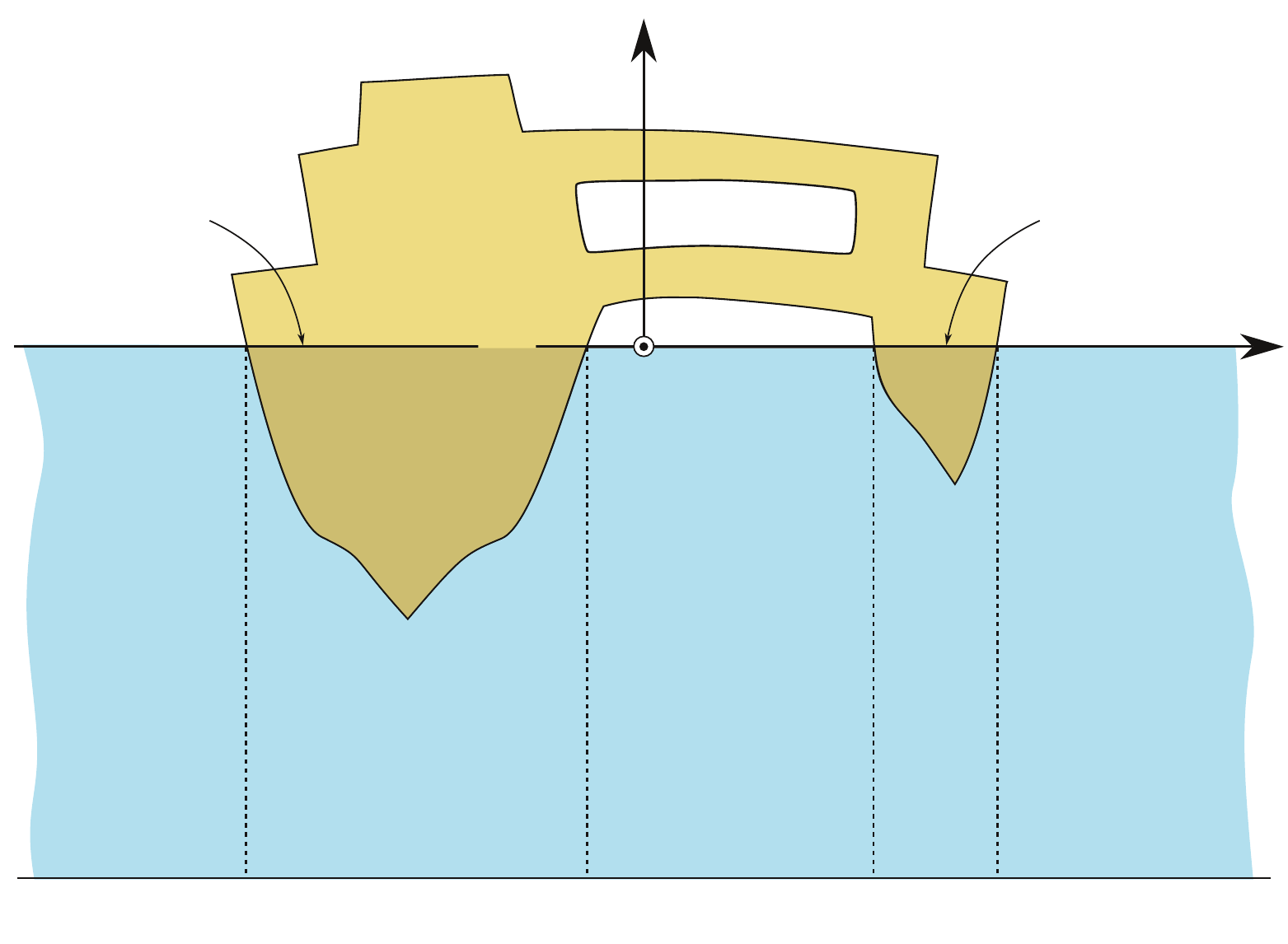}}
\end{center}
\vspace{-4mm} \caption{A sketch of the water domain $\Omega$ and the floating
surface-piercing body $\mathcal{B}$; their cross-sections by a plane orthogonal to
the $x_1$-axis are plotted. Other cross-sections marked in the figure are as
follows: $F$ denotes the free surface, whereas $B$, $\Xi$ and $\Gamma$ stand for the
bottom, the immersed part of $\mathcal{B}$ and the wetted surface of $\mathcal{B}$,
respectively; finally, $D$ is the part of $\vec{x}$-plane located within
$\mathcal{B}$.} \vspace{-4mm}
\end{figure}
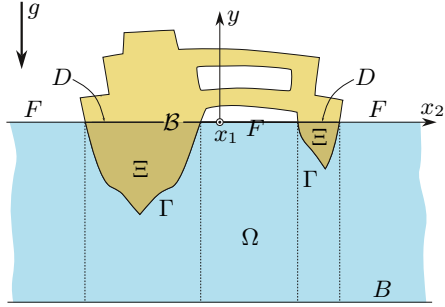

\subsection{Statement of the problem}

A Cartesian coordinate system $(x_1,x_2,y)$ (for brevity we put $(x_1,x_2) =
\vec{x}$) is chosen so that the $y$-axis is directed upwards
(gravity acts in the opposite direction), whereas the free surface of water lies
within the horizontal $\vec{x}$-plane. Let $\mathcal{B} \subset \RR^3$ denote the
domain whose closure is the floating body in its equilibrium position (see Fig.~1),
and let $\Xi$ be the non-empty submerged part of $\mathcal{B}$, that is, $\Xi =
\mathcal{B} \cap \Pi$, where
\[ \Pi = \{ (\vec{x},y) : \vec{x} \in \RR^2 , -h < y < 0 \} .
\]
We suppose that the water domain $\Omega = \Pi \setminus \overline {\mathcal B}$ is
simply connected and $\Xi$ consists of a finite number of domains adjacent to the
$\vec{x}$-plane, whereas the same number of two-dimensional domains constitutes
$D$\,---\,the interior of $\partial \Xi$ within $\{ y\!=\!0 \}$. Furthermore,
$\Gamma = \partial \mathcal B \cap \Pi$\,---\,the wetted part of body's
boundary\,---\,is supposed to form a $C^2$-surface together with the closure of its
reflection in the plane $\{ y\!=\!0 \}$. The latter condition (it guarantees that
the integral operator arising in our considerations is compact) was imposed in
\cite{J} along with John's condition. Hence, normal is defined throughout $\overline
\Gamma$ and $\vec{n}_P$ denotes the unit normal at $P = (\vec{x},y) \in \Gamma$
pointing to the interior of $\Omega$. Finally,
\[ B = \{ \vec{x} \in \RR^2 , y = -h \} \quad \mbox{and} \quad F =
\partial \Omega \setminus \overline \Gamma
\]
stand for the bottom and the free surface, respectively (see Fig.~1).

Assuming the water motion to be irrotational in $\Omega$, one concludes that it is
described by a velocity potential because the domain is simply connected. Moreover,
in the case of time-harmonic oscillations, the potential can be written as follows:
\[ \Re \bigl\{ \E^{-\ii \omega t} \varphi (\vec{x},y) \bigr\} .
\]
Here $\omega > 0$ is the radian frequency of oscillations and the complex-valued
function $\varphi$ must satisfy the following boundary value problem:
\begin{gather}
 \nabla^2 \varphi = 0 \quad \mbox{in} \ \Omega ; \label{1} \\
 \partial_y \varphi - \lambda \varphi = 0 \quad \mbox{on} \ F ; \label{2} \\
 \partial_y \varphi = 0 \quad \mbox{on} \ B ; \label{3} \\
 \partial_{\vec{n}} \varphi = v_n \quad \mbox{on} \ \Gamma ; \label{4} \\
 \int_{\Omega \cap\{|\vec{x}|=a\}} \bigl| \partial_{|\vec{x}|} \varphi - \ii k_0
 \varphi \bigr|^2 \, \D{}s = o (1) \quad \mbox{as} \quad a \to \infty . \label{5}
\end{gather}
In \eqref{1}, the Laplacian is written in terms of $\nabla = (\partial_{x_1} ,
\partial_{x_2} , \partial_y)$. The parameter $\lambda$ in \eqref{2} is equal to
$\omega^2 / g$ and attains any positive value ($g$ is the acceleration due to
gravity); $k_0$ in \eqref{5} is the unique positive root of $k \tanh k h = \lambda$.
In \eqref{4}, the given right-hand side term $v_n$ is an arbitrary continuous
function on $\overline \Gamma$ (the velocity of the body's surface in the direction
of $\vec{n}$).
 
It is natural to assume that $\varphi \in H^1_{loc} (\Omega)$, thus understanding
relations \eqref{1}--\eqref{4} in the sense of the integral identity
\begin{equation}
 \int_\Omega \nabla \varphi \nabla \psi \, \D \vec{x} \D y = \lambda \int_{F} 
 \varphi \, \psi \, \D \vec{x} - \int_\Gamma \psi \, v_n \, \D S \, ,
\label{eq:intid}
\end{equation}
which must hold for an arbitrary $\psi$ that is smooth and has a compact support in
$\overline \Omega$. It is clear that \eqref{eq:intid} follows from relations
\eqref{1}--\eqref{4} by virtue of the first Green's identity applied to $\varphi$
and $\psi$.

It is known (see, for example, \cite{MR}) that if $\varphi \in H^1_{loc} (\Omega)$
satisfies \eqref{eq:intid}, then it is continuous throughout $\overline \Omega$, and
its gradient is bounded near $\overline \Gamma \cap \overline F$. Moreover, the
Laplace equation holds in the classical sense (see, for example, \cite{GT}, ch. 8),
and the same concerns the boundary conditions \eqref{2}--\eqref{4}.

\subsection{Green's function}

Green's function $G (P, Q)$ is defined in \cite{J}, \S\,3, as a function of $P$
harmonic in $\Pi \setminus \{ Q \}$, having a source singularity at $ Q =
(\vec{\xi}, \eta) \in \Pi $ and satisfying boundary conditions on $\partial \Pi$,
namely:
\begin{itemize}

\item $\nabla^2 (G - R^{-1}) = 0$ in $\Pi$, where $R = [ r^2 + (y - \eta)^2 ]^{1/2}$
is the distance between $P$ and $Q$, whereas $r = [ (x_1 - \xi_1)^2 + (x_ 2 -
\xi_2)^2 ]^{1/2}$ is the distance between the projections of these points on a
horizontal plane;

\item $(\partial_y G - \lambda G)_{y=0} = 0$ and $(\partial_y G)_{y=-h} = 0$.

\end{itemize}
In \cite{J}, Appendix I, it is shown that there exist functions which satisfy all
these conditions and an explicit expression for a particular such function is given.
This special Green's function (denoted by $G_\lambda$ in what follows) has the
following extra properties:
\begin{itemize}

\item it is symmetric, that is, $G_\lambda (P, Q) = G_\lambda (Q, P)$ ;

\item $G_\lambda (P, Q)$ has cylindrical symmetry as well, that is, it depends only
on $y$, $\eta$ and $r$ ;

\item the radiation condition \eqref{5} holds for $G_\lambda$ as $r \to \infty$.

\end{itemize}
The last property implies the uniqueness of $G_\lambda$.

Finally, it should be mentioned that there exists an extension of $G_\lambda (P, Q)$
to the strip, where $0 < y, \eta < h$ (see \cite{J}, pp.~96, 97). For our purpose it
is sufficient to describe this extension as follows:
\begin{equation}
G_\lambda (P, Q) = R^{-1} + R_0^{-1} + H_\lambda (P, Q) \quad \mbox{for} \ -h < y,
\eta < h , \ R \neq 0 , \ R_0 \neq 0 . \label{A21}
\end{equation}
Here $R_0 = [ r^2 + (y + \eta)^2 ]^{1/2}$ is the distance from $P$ to the reflection
of $Q$ in the plane $\{ y\!=\!0 \}$, whereas $H_\lambda$ is such that for $y + \eta
\leq 0$ we have
\begin{equation}
H_\lambda (P, Q) = O (\log R_0) \ \ \mbox{and} \ \ \partial_r H_\lambda , \,
\partial_y H_\lambda , \, \partial_\eta H_\lambda = O \big( R_0^{-1} \big) \ \
\mbox{as} \ R_0 \to 0 .
\label{A22}
\end{equation}

\subsection{Irregular frequencies}

Since $G_\lambda (P, Q)$ satisfies: (1) the Laplace equation in $\Pi \setminus
\{Q\}$; (2) boundary conditions of the same kind as \eqref{2} and \eqref{3} on the
upper and lower sides of $\Pi$, respectively; (3) the radiation condition, it is
convenient to seek a solution of problem \eqref{1}--\eqref{5} in the form of a
single-layer potential
\[ (V \mu) (P) = \int_\Gamma \mu (Q) \, G_\lambda (P,Q) \, \D S_Q \quad \mbox{with an
unknown} \ \mu \in C (\overline \Gamma) \, .
\]
Here $C (\overline \Gamma)$ is the Banach space of continuous complex-valued
functions on $\overline \Gamma$. Indeed, if one finds $\mu$ from the Neumann
boundary condition \eqref{4}, then $V \mu$ is the problem's solution. The theorem
about the limit values of the normal derivative of $V \mu$ on $\Gamma$ (see, for
example, \cite{M}, Ch.~18, \S\,7) yields the integral equation:
\begin{equation}
- \mu (P) + \frac{1}{2\pi} \int_\Gamma \mu (Q) \, \frac{\partial G_\lambda}
{\partial n_P} (P,Q) \, \D S_Q = \frac{1}{2\pi} v_n (P) , \quad P \in \Gamma ;
\label{ie}
\end{equation}
$(-I + K') \, \mu = f$ for short, where $I$ is the identity operator and $f = (2
\pi)^{-1} v_n$.

Following \cite{CK}, Ch.~2, let us consider this equation in the framework of the
dual system $\langle C (\overline \Gamma), C (\overline \Gamma) \rangle$ equipped
with the bilinear form
\begin{equation}
\langle \mu , \nu \rangle = \int_\Gamma \mu \, \nu \, \D S .
\label{du}
\end{equation}
The integral operator $K'$ in \eqref{ie} is adjoint to
\[ (K \nu) (P) = \frac{1}{2\pi} \int_\Gamma \nu (Q) \, \frac{\partial G_\lambda}
{\partial n_Q} (P,Q) \, \D S_Q 
\]
with respect to the bilinear form $\langle \cdot , \cdot \rangle$. Here and below
the notation for operators adopted in \cite{CK} is used.

Let us show that $K'$ (and $K$ as well) is a compact operator in $C (\overline
\Gamma)$. Formula \eqref{A21} for $G_\lambda$ allows us to represent this operator
as follows:
\begin{equation}
(K' \mu) (P) = \frac{1}{2\pi} \left[ \int_{\Gamma_0} \hat \mu (Q) \,
\frac{\partial R^{-1}} {\partial n_P} (P,Q) \, \D S_Q + \int_\Gamma \mu (Q) \,
\frac{\partial H_\lambda} {\partial n_P} (P,Q) \, \D S_Q \right] .
\label{K'}
\end{equation}
Here $\Gamma_0$ is the union of $\overline \Gamma$ and its reflection in the plane
$\{ y\!=\!0 \}$. By $\hat \mu$ we denote the function equal to $\mu$ on $\overline
\Gamma$, whereas on the upper half of the doubled surface $\hat \mu$ is the
extension of $\mu$ even with respect to~$y$. Since $\Gamma_0$ is a $C^2$-surface,
the kernel $\partial R^{-1} / \partial n_P (P,Q)$ is weakly singular, and so the
operator corresponding to the first term in square brackets is compact (see
\cite{CK}, \S\S\,2.3 and 2.7). The operator defined by the second term in square
brackets is compact in view of the second set of estimates \eqref{A22}.

Thus, \eqref{ie} is a Fredholm equation; it is analogous to
\begin{equation}
(-I + K_0') \, \mu = f \quad \mbox{with} \ \ (K'_0 \mu) (P) = \frac{1}{2\pi}
\int_{\Gamma_0} \mu (Q) \, \frac{\partial R^{-1}} {\partial n_P} (P,Q) \, \D S_Q \,
, \label{K'_0}
\end{equation}
arising when one applies the potential theory to the Neumann problem for the Laplace
equation in the domain exterior to $\Gamma_0$ (see \cite{M}, Ch.~18, \S\S\,8 and
10). It should be emphasized that $K'_0$ is not the limit of $K'$ as $\lambda \to 0$
(see \cite{J}, p.~97), the subscript in $K'_0$ just means that integration is over
$\Gamma_0$. There is an essential distinction between \eqref{K'_0} and \eqref{ie}
because the homogeneous equation \eqref{K'_0} has only a trivial solution, whereas
the dependence of $K'$ on $\lambda$ results in the existence of the so-called
irregular frequencies for which the homogeneous equation \eqref{ie} has nontrivial
solutions.

Namely, if for some $\lambda_*$ the homogeneous equation \eqref{ie} has a nontrivial
solution $\mu_*$, then this value of the parameter is called irregular as well as
the corresponding value of frequency. In the {\it irregular case} (this notion was
introduced by John; see \cite{J}, p.~85), John's condition guarantees that $V \mu_*
\equiv 0$ on $\overline \Omega$, whereas $V \mu_*$ is nontrivial at least in
$\Xi_k$\,---\,one of connected components of $\Pi \setminus \overline \Omega$ (see
Fig.~1, where two such components are shown). Indeed, there exists an eigenfunction
of the following spectral problem
\[ \nabla^2 \psi = 0 \ \mbox{in} \ \Xi_k , \quad \partial_y
\psi - \lambda \psi = 0 \ \mbox{on} \ D_k , \quad \psi = 0 \ \mbox{on} \ \Gamma_k ,
\]
corresponding to the eigenvalue $\lambda = \lambda_*$. Moreover, the fact that
$\lambda$ is an eigenvalue of this problem for some $\Xi_k$ is a necessary and
sufficient condition that $\lambda$ is irregular (for the proof see, for example,
\cite{LWW}, p.~105).

It is known that the assumptions made about $\Gamma$ yield that for every $\Xi_k$
there exists a sequence
\[ 0 < \lambda_1^{(k)} \leq \dots \leq \lambda_n^{(k)} \leq \dots
\ \mbox{such that} \ \lambda_n^{(k)} = \sqrt{\frac{|D_k|}{4 \pi n}} \, [1 + o (1)] \
\mbox{as} \ n \to \infty ;
\]
here $|D_k|$ denotes the area of $D_k$. Thus, the whole set of irregular values for
equation \eqref{ie} is the union of the finite number of these sequences.

\subsection{Background on equations without irregular frequencies}

The question how to find a solution of problem \eqref{1}--\eqref{5} in the case when
$\lambda$ is irregular has a long history; see \cite{LWW}, \S\,3.1.2, for a survey
of various approaches; a few of them are characterized below.

The idea of the first method is to replace equation \eqref{ie} by an
integro-algebraic system; see \cite{J}. The latter consists of \eqref{ie} amended by
adding a linear combination of $m$ given functions with unknown coefficients, and a
linear algebraic system for these coefficients. Here $m$ is the number of linearly
independent solutions of the homogeneous equation \eqref{ie}, whereas the functions
are constructed on the basis of these solutions. In \cite{NK}, this approach was
extended to the case of piecewise smooth $\Gamma$ with edges and vertices.

A rather simple method developed by Ursell \cite{U} consists in modifying Green's
function in order to obtain an integral equation without irregular frequencies. This
idea was brought to the theory of time-harmonic water waves from acoustics, where it
had demonstrated its fruitfulness. Another approach borrowed from acoustics is to
seek a solution as the sum of $V \mu$ and the volume Green's potential over $\Xi$
which leads to a system of two integral equations; see details in \cite{LWW},
\S\,3.1.1.3, where the presentation follows \cite{NGK}.


\subsection{Main result}

The aim of this paper is to show that there is a single equation with operators
involving only $G_\lambda$ and such that no irregular values of $\lambda$ exists for
it; that is, this equation is uniquely soluble for all right-hand side terms. For
this purpose let us seek $\varphi$ in the form
\begin{equation}
(V \mu) (P) + (W \nu) (P) , \label{VW}
\end{equation}
where $V \mu$ is the simple-layer potential with unknown density, whereas
\[ (W \nu) (P) = \int_\Gamma \nu (Q) \, \frac{\partial G_\lambda}{\partial n_Q} 
(P, Q) \, \D S_Q 
\]
is the double-layer potential whose density is also unknown. It is clear that this
representation with $\mu, \, \nu \in C (\overline \Gamma)$ yields that $\varphi$
satisfies all relations of the boundary value problem except for the Neumann
condition on~$\Gamma$. However, the continuity of $\nu$ is insufficient for the
existence of $\partial_{\vec{n}_P} (W \nu)$ on $\Gamma$. This fact is attributed to
Lyapunov in \cite{G}, Ch.~II, \S\,9, and elsewhere, but it is just mentioned in the
note \cite{L}, whereas an example of the surface and density demonstrating this fact
for
\[ (W_0 \nu) (P) = \int_{\Gamma_0} \nu (Q) \, \frac{\partial R^{-1}}
{\partial n_Q} (P,Q) \, \D S_Q 
\]
first appeared in Steklov's monograph \cite{S}, \S\,34. It is worth mentioning
that the main result of \cite{L} is a condition expressed in terms of local
properties of $\nu$ near $P$ that guarantees the existence of $\partial_{\vec{n}_P}
(W_0 \nu)$ at $P \in \Gamma_0$. It is not known how close is this condition to a
necessary one.

Two decades ago, Lukyanov and Nazarov \cite{LN} obtained a general criterion for the
global existence of the normal derivative of $W_0 \nu$ (of course, their criterion
does not require symmetry of $\Gamma_0$). Combining this criterion with \eqref{A21}
and \eqref{A22}, one obtains that for the global existence of $\partial_{\vec{n}_P}
(W \nu)$ on $\overline \Gamma$ it is necessary and sufficient that there exists
$\nu_* \in C (\overline \Gamma)$ such that $\nu = [V \nu_*]_{\overline \Gamma}$. The
set $\mathcal D$ consisting of densities representable as traces of single-layer
potentials is a non-empty linear variety in $C (\overline \Gamma)$. Indeed, a
density belongs to $\mathcal D$ provided it is in $C^{1, \alpha} (\overline \Gamma)$
with a certain exponent $\alpha \in (0, 1)$; see \cite{CK}, \S\,2.7. Moreover, every
$\nu \in \mathcal D$ is a H\"older continuous function on $\overline \Gamma$ with
$\alpha \in (0, 1)$, and so $W \nu$ is uniformly H\"older continuous on both
$\overline \Omega$ and $\overline \Xi$ with $\alpha \in (0, 1)$; moreover, the
H\"older norms of $W \nu$ on these closed sets are estimated by that of $\nu$ (see
\cite{CK}, \S\,2.4).

Seeking $\varphi$ in the form \eqref{VW} with unknown $\mu \in C (\overline \Gamma)$
and $\nu \in \mathcal D$, we obtain the equation
\begin{eqnarray}
&& - \mu (P) + \frac{1}{2\pi} \int_\Gamma \mu (Q)\, \frac{\partial G_\lambda}
{\partial n_P} (P,Q) \, \D S_Q \nonumber \\ && \ \ \ \ \ \ \ \ \ \, + \frac{1}{2\pi}
\frac{\partial}{\partial n_P} \int_\Gamma \nu (Q) \, \frac{\partial G_\lambda}
{\partial n_Q} (P,Q) \, \D S_Q = \frac{1}{2\pi} v_n (P) \, , \quad P \in \Gamma 
\label{ine}
\end{eqnarray}
from the boundary condition \eqref{4}. In what follows, we write 
\[ (-I + K') \, \mu + T \, \nu = f
\]
for short, where $T$ denotes the integro-differential operator in the second line of
\eqref{ine}. It is worth mentioning that it was Kleinman \cite{Kl} who first used
the operator $T$ in his treatment of the floating-body problem. In his approach,
this operator arose in the framework of a direct version of boundary integral
technique.

Now we are in a position to formulate the following assertions.

\begin{theorem}
Let the surface $\Gamma$ satisfy John's condition and let the union of\ $\overline
\Gamma$ and its reflection in the plane $\{ y\!=\!0 \}$ be a $C^2$-surface, then for
all $\lambda > 0$ the indefinite equation \eqref{ine} with an arbitrary $v_n \in C
(\overline \Gamma)$ has a solution $(\mu, \nu)$, where $\mu \in C (\overline
\Gamma)$ and $\nu \in \mathcal D$.
\end{theorem}

\begin{corollary}
Let the assumptions imposed on $\Gamma$ in Theorem~1 hold, then for all $\lambda >
0$ problem \eqref{1}--\eqref{5} with an arbitrary $v_n \in C (\overline \Gamma)$ has
a solution of the form \eqref{VW}, where $\mu \in C (\overline \Gamma)$ and $\nu \in
\mathcal D$.
\end{corollary}


\section{Proof of Theorem 1 and Corollary 1}

\subsection{Self-adjointness of $\mathbf T$}

Our proof of Theorem 1 is based on Theorem~2 (see Appendix), and the latter theorem
involves the operator $T'$ defined in the dual system $\langle C (\overline \Gamma)
, C (\overline \Gamma) \rangle$ (see \eqref{du} for the duality relation). The
situation is simplified by the fact that $T$, which is unbounded in $C (\overline
\Gamma)$, has the self-adjoint closure being symmetric on $\mathcal D$. Indeed,
$\langle T \nu_1 , \nu_2 \rangle = \langle \nu_1 , T \nu_2 \rangle$ for all $\nu_1 ,
\nu_2 \in \mathcal D$. This fact was earlier established in the case of a similar
operator in which the fundamental solution of the Helmholtz equation used as the
kernel (see \cite{CK}, \S\,2.7). Therefore, we just outline the proof in our case.

It is sufficient to suppose that $\nu_k$, $k=1,2$, are in $C^{1, \alpha} (\overline
\Gamma)$, $\alpha \in (0, 1)$, in which case the first derivatives of $W \nu_k$ are
uniformly extendable from $\Omega$ ($\Xi$) to $\overline \Omega$ ($\overline \Xi$
respectively) as H\"older continuous functions (see \cite{CK}, \S\,2.5).

Let $u_k = W \nu_k$, then the jump relation for this potential yields
\[ \langle T \nu_1 , \nu_2 \rangle = \int_\Gamma (T \nu_1) \, \nu_2 \, \D S = 2 
\int_\Gamma \frac{\partial u_1}{\partial n} \, \big( u_2^\Omega - u_2^\Xi \big) \,
\D S \, ,
\]
where $u_2^\Omega$ ($u_2^\Xi$) is the limiting value of $u_2$, when approaching
$\Gamma$ from $\Omega$ ($\Xi$ respectively). The second Green's formula allows us to
continue this chain of equalities
\[ 2 \int_\Gamma \big( u_1^\Omega - u_1^\Xi \big) \, \frac{\partial u_2}{\partial n} 
\, \D S = \int_\Gamma \nu_1 \, T \nu_2 \, \D S = \langle \nu_1 , T \nu_2 \rangle .
\]

Another property to be used is that $K$ and $K'$ are compact operators in $C
(\overline \Gamma)$. Moreover, in the same way as in \cite{CK}, \S\,2.7, one obtains
that these operators map $C (\overline \Gamma)$ to $C^{0, \alpha} (\overline
\Gamma)$, $\alpha \in (0, 1)$, whereas $K$ maps $C^{0, \alpha} (\overline \Gamma)$
to $C^{1, \alpha} (\overline \Gamma)$.

\subsection{Proof of Theorem 1}

According to Theorem 2, equation \eqref{ine} is soluble for all right-hand side
terms, if and only if the intersection of the null-spaces of $-I+K$ and $T$ is
trivial.

To show this we consider $\mu_0 \in C (\overline \Gamma)$ satisfying
\begin{equation}
(-I+K) \mu_0 = {\bf 0} , \label{0}
\end{equation}
where ${\bf 0}$ is the zero element of $C (\overline \Gamma)$. Then the properties
of $K$ mentioned at the end of \S\,2.1 imply that $\mu_0 \in C^{1, \alpha}
(\overline \Gamma)$, and so $T \mu_0$ is well defined. If $T \mu_0 = {\bf 0}$ (that
is, $\mu_0$ belongs to the intersection of the null-spaces of $-I+K$ and $T$), then
$W \mu_0$ solves problem \eqref{1}--\eqref{5} with $v_n \equiv 0$ on $\Gamma$.

Since John's condition guarantees the uniqueness of solution for this problem, we
have that $(W \mu_0) (P) = 0$ for all $P \in \Omega$. Letting $P \to \Gamma$, we
obtain that $(K+I) \mu_0 = {\bf 0}$ by the theorem about limiting values of the
double-layer potential. Combining the last equality and \eqref{0}, we conclude that
$\mu_0 = {\bf 0}$, which completes the proof.

\subsection{Proof of Corollary 1}

Seeking a solution of problem \eqref{1}--\eqref{5} as the sum of two potentials
\eqref{VW}, one arrives at the indefinite equation \eqref{ine} for the unknown
densities $\mu$ and $\nu$. According to Theorem~1, this equation is soluble for
every continuous right-hand side term. Substituting a solution of \eqref{ine} into
\eqref{VW}, one obtains that of problem \eqref{1}--\eqref{5} which is unique in view
of John's condition imposed on the geometry of $\Gamma$. The proof is complete.

\subsection{Discussion}

The solubility of problem \eqref{1}--\eqref{5} is established via reducing it to the
indefinite equation \eqref{ine} under a rather restrictive assumption about
smoothness of the symmetric surface $\Gamma_0$ which appears in formula \eqref{K'}.
The assumption that $\Gamma_0$ is a $C^2$-surface is imposed in order to refer
directly to assertions proved in \cite{CK}, where this class of surfaces is
considered. It is straightforward to modify our proofs to the case when $\Gamma_0$
is a $C^{1, \alpha}$-surface, $\alpha \in (0, 1)$ (this class is also referred to as
Lyapunov surfaces).

Meanwhile, it is known that the single- and double-layer potentials for the Laplace
equation are well-defined on surfaces enclosing bounded Lipschitz domains. Using
these potentials, Verchota \cite{V} reduced the interior Dirichlet and Neumann
problems to integral equations with operators analogous to $I+K$ and $I-K'$
respectively. Moreover, he demonstrated the invertibility of the first of these
operators in $L^2$, whereas the second one is invertible in the subspace of $L^2$
orthogonal to constants. This allowed him to show that the Dirichlet and Neumann
problems are soluble and their solutions are representable as the corresponding
potentials.

The results obtained in \cite{V} can serve as an initial step towards generalizing
the approach to problem \eqref{1}--\eqref{5} based on the indefinite equation
\eqref{ine} in the case when $\Gamma_0$ is a Lipschitz surface satisfying John's
condition. However, there are several points that make treatment of this equation
more complicated in the latter case. First, the equation must be considered in the
dual system $\langle H^{1/2} (\Gamma), H^{-1/2} (\Gamma) \rangle$ with the duality
relation between these Sobolev spaces extending \eqref{du}. Indeed, the operator $T$
arising in this case maps continuously $H^{1/2} (\Gamma)$ to $H^{-1/2} (\Gamma)$
and, presumably, is self-adjoint in this dual system; see considerations in
\cite{Kr}, \S\,8.3. Another point is to check whether $K$ and $K'$ are compact in
the mentioned dual system. Finally, it must be shown that Theorem~2 is true in this
system.

\section*{Appendix: Theorem on solubility \\ of indefinite equations}

We consider the question of solubility of the following indefinite equation
\[ L u + M v = f 
\]
in the framework of a dual system $\langle X, X \rangle$, where $X$ is a Banach
space. Here, $L$ and $M$ are linear operators acting from $X$ to $X$, $f \in X$ is a
given element, whereas $u$ and $v$ are unknown elements of $X$ to be found from the
equation.

We recall that a normed space $X$ equipped with a non-degenerate bilinear form
$\langle \cdot, \cdot \rangle : X \times X \to \CC$ is a dual system (generally
speaking, two normed spaces are used in the definition of a dual system; see, for
example \cite{CK}, \S\,1.3). For $M$, which can be unbounded, but has a domain
$\mathcal{D}_M$ dense in $X$, there exists $M'$\,---\,the operator adjoint to $M$
with respect to $\langle \cdot, \cdot \rangle$\,---\,provided there is a domain
$\mathcal{D}_{M'} \subset X$ and the mapping $\mathcal{D}_{M'} \ni v' \mapsto M' v'
\in X$ such that
\begin{equation} 
\langle M v, v' \rangle = \langle v, M' v' \rangle \quad \mbox{for all} \ v \in
\mathcal{D}_M \ \mbox{and} \ v' \in \mathcal{D}_{M'} . \label{adj}
\end{equation}
To distinguish the adjoint operator in the dual system $\langle X, X \rangle$, we
mark it with~$'$ instead of $^*$ denoting the standard adjoint operator.

In what follows, $\mathcal R (M)$ and $\mathcal N (M)$ denote the range and the null
space, respectively, for an operator $M$; $I$ stands for the identity operator and
${\bf 0}$ is the zero element of $X$.

\begin{theorem}
Let $\langle X, X \rangle$ be a dual system on a Banach space $X$. Let $C$ be a
compact operator mapping $X$ into itself and let $M$ be a closed operator from $X$
to $X$ having $\mathcal{D}_M$ dense in $X$. The indefinite equation
\begin{equation} 
(I + C) u + M v = f  \label{eq}
\end{equation}
has a solution for all $f \in X$ if and only if $\mathcal N (I + C') \cap \mathcal N
(M') = {\bf 0}$. Here $M'$ is defined by \eqref{adj} and $C'$ is defined similarly.

If $u$ and $v$ satisfy \eqref{eq} and are such that
\begin{equation} 
u \in \mathcal R (I + C') , \quad v \in M' [ \mathcal N (I + C') ] , \label{cond}
\end{equation}
then the solution $(u, v)$ is unique in $\mathcal R (I + C') \times M' [ \mathcal N
(I + C') ]$.
\end{theorem}

This theorem is a slight improvement of a theorem of S. G. Krein \cite{K}; see also
his book \cite{SK}, \S\,18. The proof is given to make the paper self-contained.

\begin{proof}
$1^\circ$. We begin with proving the second assertion, namely, conditions
\eqref{cond} guarantee that the homogeneous equation
\begin{equation} 
(I + C) u_0 + M v_0 = {\bf 0} \label{heq}
\end{equation}
has only a trivial solution $(u_0, v_0) = ({\bf 0}, {\bf 0})$. According to the
second condition \eqref{cond}, there exists $v' \in \mathcal N (I + C')$ such that
$v_0 = M' v'$. Since 
\[ \langle (I + C) u_0, v' \rangle = \langle u_0, (I + C') v' \rangle = 0 ,
\]
equation \eqref{heq} yields that
\[ 0 = \langle {\bf 0}, v' \rangle = \langle (I + C) u_0, v' \rangle + \langle M v_0, 
v' \rangle = \langle v_0, M' v' \rangle = \langle v_0, v_0 \rangle \, ,
\]
and so $v_0 = {\bf 0}$. Now, \eqref{heq} takes the form $(I + C) u_0 = {\bf 0}$,
where $u_0 = (I + C') u'$ by the first condition \eqref{cond}. Hence we have
\[ 0 = \langle {\bf 0}, u' \rangle = \langle (I + C) u_0, u' \rangle = \langle u_0, 
(I + C') u' \rangle = \langle (I + C') u', (I + C') u' \rangle \, ,
\]
that is, $u_0 = (I + C') u' = {\bf 0}$, which completes the proof of the second
assertion.

\vspace{2mm}

$2^\circ$. By $\mathcal F$ we denote the set of those $f$, for which \eqref{eq} has
a solution satisfying conditions \eqref{cond}. Let us show that $\mathcal F$ is
closed in $X$. For this purpose we consider a sequence $\{ f_n \} \subset \mathcal
F$ with the following properties: $f_n \to f$ as $n \to \infty$, and there exist $\{
u_n \} \subset \mathcal R (I + C')$ and $\{ v_n \} \subset M' [ \mathcal N (I + C')
]$ such that
\begin{equation} 
(I + C) u_n + M v_n = f_n \quad \mbox{for} \ n = 1,2,\dots \, .  \label{eq_n}
\end{equation}
Let us demonstrate that equation \eqref{eq} is soluble for $f$ and its solution is
such that $u \in \mathcal R (I + C')$ and $v \in M' [ \mathcal N (I + C') ]$.

First, let  $\{ u_n \}$ and $\{ v_n \}$ be bounded sequences. Then there exists a
subsequence $\{ n' \}$ such that $C u_{n'}$ has a limit as $n' \to \infty$ because
$C$ is compact. Since $M' [ \mathcal N (I + C') ]$ is finite-dimensional and $M$ is
a closed operator, there is a subsequence $\{ n'' \}$ of $\{ n' \}$ and $v \in M' [
\mathcal N (I + C') ]$ such that
\[ v_{n''} \to v \ \ \mbox{and} \ \ M v_{n''} \to M v \in M' [ \mathcal N (I + C') ]
\quad \mbox{as} \ n'' \to \infty \, .
\]
Letting $n = n'' \to \infty$ in \eqref{eq_n}, we see that $u$ and $v$ solve
\eqref{eq}. Since $C'$ is compact along with $C$, we have that $\mathcal R (I + C')$
is closed, and so $u$ belongs to this set.

The same considerations remain valid in the case when $\{ u_n \}$ is unbounded, that
is, there exists a subsequence $\{ n' \}$ such that $ \|u_{n'}\|$ tends to infinity
as $n' \to \infty$, but the sequence $\{ \|v_{n'}\| / \|u_{n'}\| \}$ is bounded.

$3^\circ$. To complete the proof that $\mathcal F$ is closed we have to consider the
case when $\{ u_n \}$ is unbounded and there exists a subsequence $\{ n'' \}$ such
that
\[ \|v_{n''}\| / \|u_{n''}\| \to \infty \quad \mbox{as} \ n'' \to \infty .
\]
 Then
\[ (I + C) \frac{u_{n''}}{\|v_{n''}\|} + M \frac{v_{n''}}{\|v_{n''}\|} = 
\frac{f_{n''}}{\|v_{n''}\|} \to {\bf 0} \quad \mbox{as} \ n'' \to \infty \, ,
\]
which implies that $M v = {\bf 0}$ for some $v \in M' [ \mathcal N (I + C') ]$ with
$\|v\| = 1$, which is incompatible with the second assertion of the theorem. Thus,
we conclude that the sequence $\{ \|u_n\| \}$ is bounded. In the same way, one
obtains that $\{ \|v_n\| \}$ is bounded.

$4^\circ$. It is clear that $X \neq \mathcal F$ if and only if there is $f_* \neq
{\bf 0}$ such that
\begin{equation} 
\langle (I + C) u , f_* \rangle + \langle M v, f_* \rangle = 0 \ \mbox{for all}
\ u \in \mathcal R (I + C') \ \mbox{and} \ v \in M' [ \mathcal N (I + C') ] \, .
\label{eq_*}
\end{equation}
For $v = {\bf 0}$ this yields
\[ \langle u , (I + C') f_* \rangle = 0 \quad \mbox{for all} \ u \in \mathcal R 
(I + C') \, ,
\]
and so for $u = (I + C') f_*$ we obtain
\begin{equation} 
(I + C') f_* = {\bf 0} , \quad \mbox{that is}, \ f_* \in \mathcal N (I + C') \, .
\label{f_*1}
\end{equation}
Substituting $u = {\bf 0}$ into \eqref{eq_*}, we see that 
\[ \langle M v, f_* \rangle = \langle v, M' f_* \rangle = 0 \quad \mbox{for all} \ v 
\in M' [ \mathcal N (I + C') ] \, .
\]
For $v = M' f_*$ this gives
\begin{equation} 
M' f_* = {\bf 0} , \quad \mbox{that is}, \ f_* \in \mathcal N (M') \, .
\label{f_*2}
\end{equation}
Thus, \eqref{f_*1}  and \eqref{f_*2} imply that $X = \mathcal F$ if and only if
$\mathcal N (I + C') \cap \mathcal N (M') = {\bf 0}$. The proof of the theorem's
first assertion is complete.
\end{proof}


{\small
\begin{thebibliography}{99}

\bibitem{CK} Colton, D., Kress, R., \emph{Integral Equation Methods in Scattering
Theory}, SIAM, Philadelphia, 2013.

\bibitem{GT} Gilbarg, D., Trudinger, N. S., \emph{Elliptic Partial Differential
Equations of Second Order}, Springer, Berlin et al. 1983.

\bibitem{G} G\"unther, N. M., \emph{Potential Theory and Its Applications to Basic
Problems of Mathematical Physics}. Ungar, New York, 1967.

\bibitem{J} John, F., \emph{On the motion of floating bodies. II}, Comm. Pure Appl.
Math. {\bf 3} (1950), 45--101.

\bibitem{Kl} Kleinman, R. E., \emph{On the mathematical theory of motion of floating
bodies\,--\,an update}, D. W. Taylor Naval Ship Res. \& Devel. Center, Report
82/074 (1982).

\bibitem{K} Krein, S. G., \emph{On one undetermined equation in Hilbert space and
its application to potential theory}, Uspekhi Mat. Nauk {\bf 9}, no. 3 (1954),
149--153 (in Russian).

\bibitem{SK} Krein, S. G., \emph{Linear Equations in Banach Spaces}, Birkh\"auser,
Boston, Basel, Stuttgart, 1982.

\bibitem{Kr} Kress, R., \emph{Linear Integral Equations}, Springer, New York, 2014.

\bibitem{NK} Kuznetsov, N. G., \emph{Steady waves on the surface of a water layer of
variable depth with immersed floating bodies}, Regular Asymptotic Algorithms in
Mechanics, Nauka, Novosibirsk, 1989, pp.~200--261 and 268--270.

\bibitem{NGK} Kuznetsov, N. G., \emph{Integral equations for the problem of
stationary waves produced by a floating body}, Math. Notes {\bf 50} (1991),
1036--1042.

\bibitem{LWW} Kuznetsov, N., Maz'ya, V., Vainberg, B., \emph{Linear Water Waves: A
Mathematical Approach}, Cambridge University Press, Cambridge, 2002.

\bibitem{L} Liapounoff, A. M., \emph{Sur le potentiel de la double couche}, Comm.
Soc. math. Kharkow, S\'er. 2, {\bf 6} (1899), 129--138.

\bibitem{LN} Lukyanov, V. V., Nazarov, A. I., \emph{Solving the Venttsel’ problem
for the Laplace and Helmholtz equations with the help of iterated potentials},
Zapiski Nauch. Semin. POMI {\bf 250} (1998), 203--218. (in Russian; English transl.
J. Math. Sci. {\bf 102} (2000), 4265--4274).

\bibitem{MM} McIver, M., \emph{An example of non-uniqueness in the two-dimensional
linear water wave problem}, J.\ Fluid Mech. {\bf 315} (1996), 257--266.

\bibitem{MP} McIver, P., McIver, M., \emph{Trapped modes in an axisymmetric water
wave problem}, Quart. J. Mech. Appl. Math. {\bf 50} (1997), 165--178.

\bibitem{MR}  Maz'ya, V. G., Rossmann, J., \emph{\"Uber die Asymptotic der
L\"osungen elliptischer Randwertaufgaben in der Umgebung von Kanten}, Math. Nachr.
{\bf 138} (1988), 27--53.

\bibitem{M} Mikhlin, S. G., \emph{Mathematical Physics, an Advanced Course},
North-Holland Publishing Co., Amsterdam-London, 1970.

\bibitem{S} Steklov, V. A., \emph{Fundamental Problems of Mathematical Physics},
Vol.~2, Russian Academy of Sciences, Petrograd, 1923. (2nd ed. Nauka, Moscow, 1983;
both in Russian).

\bibitem{U} Ursell, F., \emph{Irregular frequencies and the motion of floating
bodies}, J. Fluid Mech. {\bf 105} (1981), 143--156.

\bibitem{V} Verchota, G., \emph{Layer potentials and regularity for the Dirichlet
problem for Laplace’s equation in Lipschitz domains}, J. Func. Anal. {\bf 39}
(1984), 572--611.

\end{thebibliography}
}
\end{document}